\documentclass[conference]{IEEEtran}
\IEEEoverridecommandlockouts
\usepackage{cite}
\usepackage{amsmath,amssymb,amsfonts}
\usepackage{graphicx}
\usepackage[T1]{fontenc}
\usepackage[table]{xcolor}
\usepackage{tcolorbox}
\tcbuselibrary{skins}
\usepackage{textcomp}
\usepackage{xcolor}
\usepackage{enumitem}
\usepackage{url}
\usepackage{algorithm}
\usepackage{algpseudocode}
\usepackage{tcolorbox}
\usepackage[dvipsnames]{xcolor}
\usepackage{booktabs}
\usepackage{siunitx}
\usepackage{multirow}
\usepackage{cprotect}
\usepackage{tabularx}
\usepackage{booktabs}
\usepackage{fvextra}
\usepackage{array}
\def\BibTeX{{\rm B\kern-.05em{\sc i\kern-.025em b}\kern-.08em
    T\kern-.1667em\lower.7ex\hbox{E}\kern-.125emX}}

\begin{document}

\title{Towards Energy-aware Requirements Dependency Classification: Knowledge-Graph vs. Vector-Retrieval Augmented Inference with SLMs
%
}
\author{
Shreyas Patil, Pragati Kumari, Novarun Deb, Gouri Ginde \\
\{shreyas.patil, pragati.kumari, novarun.deb, gouri.ginde\}@ucalgary.ca \\
Dept. Electrical and Software Engineering, University of Calgary, Canada
}

\maketitle

\begin{abstract}
The continuous evolution of system specifications necessitates frequent evaluation of conflicting requirements, a process that is traditionally labor-intensive. Although large language models (LLMs) have demonstrated significant potential in automating this detection, their massive computational requirements often lead to excessive energy waste. Consequently, there is a growing need to transition toward Small Language Models (SLMs) and energy-aware architectures for sustainable Requirements Engineering. This study proposes and empirically evaluates a energy-aware framework that compares Knowledge Graph-based Retrieval (KGR) with Vector-based Semantic Retrieval (VSR) to enhance SLM-based inference at the 7B–8B parameter scale. By leveraging structured graph traversal and high-dimensional semantic mapping, we extract requirement candidates which are then classified as conflicting or neutral through an inference engine. We evaluate these retrieval-enhanced strategies across Zero-Shot, Few-Shot, and Chain-of-Thought prompting methods. Using a three-pillar sustainability framework—measuring energy consumption (Wh), latency (s), and carbon emissions (gCO$_2$eq)—alongside standard accuracy metrics (F1-score),  this research provides  a first systematic empirical evaluation and trade-off analysis between predictive performance and environmental impact. Our findings highlight the effectiveness of structured versus semantic retrieval in detecting requirement conflicts, offering a reproducible, sustainability-aware architecture for energy-efficient requirement engineering.
\end{abstract}


\section{Introduction}
\label{sec:Intro}
Requirements interdependencies define how requirements relate to and influence one
another \cite{dahlstedt2005requirements}, playing a critical role in the delivery of robust software systems. Despite their importance, existing literature lacks a unified framework to systematically capture and classify these interdependencies \cite{gharib2025reinta}. \textit{Requires}, \textit{Conflict}, \textit{Similar} and \textit{Refines} are some of the common relation types. However, identification and classification of \textit{Conflict} dependency type - where fulfilling one requirement may hinder, contradict, exclude, or violate another (e.g., performance vs. security trade-offs, resource constraints, temporal/logical incompatibilities) \cite{dahlstedt2005requirements} \cite{nuseibeh2002leveraging} \cite{HowtoHan57:online} -  remains a significant and active challenge in Requirements Engineering (RE) mainly due to the subtle semantic contradictions, negations, domain-specific constraints, and multi-requirement interactions beyond basic NLP/pattern-matching \cite{cheng2026generative} \cite{saleem2025passionnet}. Unidentified or poorly managed  conflicting requirements can lead organizations to delays, wasted resources, and risks to compliance and customer satisfaction \cite{Gartner2024} \cite{HowtoHan57:online}. 

\textbf{Problem. }Recent advancement in the state-of-the-art of Large Language Models (LLMs) has transformed the landscape of RE \cite{saleem2025passionnet}, demonstrating a high potential to capture the fine semantic nuances required for tasks such as requirement traceability and specification analysis ~\cite{Niu2025, Klesel2025}. In the context of Conflict dependency type detection, where deep linguistic understanding is vital to prevent costly late-stage corrections~\cite{Sommerville2016}, these models bridge the gaps that traditional information retrieval techniques often struggle to address~\cite{Zadenoori2025, hey_requirements_2025}. However, the deployment of massive LLMs is hampered by significant computational overhead, characterized by high energy consumption, large carbon footprints, and the lack of a specific project context~\cite{Norheim2024}. This creates a pressing need to explore how Small Language Models (SLMs), typically in the 7B-8B range, perform similar classification tasks. Although SLMs are more resource-efficient, they require robust, context-sensitive augmentation to handle the full scale of complex software repositories~\cite{lewis2020retrieval}. Consequently, the transition from brute-force inference to context-aware retrieval is essential to make SLM-based RE both practical and accurate.

State-of-the-art retrieval-augmented techniques primarily rely on two paradigms: Vector-based Semantic Retrieval (\textit{VSR}) and Knowledge Graph-based Retrieval (\textit{KGR}). \textit{VSR} offers superior latency by encoding requirements in high-dimensional spaces yet it frequently suffers from \textit{structural blindness} - an inability to map the hierarchical and relational dependencies inherent in software specifications. In contrast, \textit{KGR} systems provide a structured environment that defines these subtle dependencies, which is essential for avoiding the misidentification of conflicts. Although \textit{KGR} is computationally more intensive during the initial indexing phase, evidence suggests that it may reduce long-term environmental costs by eliminating \textit{contextual noise} that forces models into redundant and energy-intensive processing cycles~\cite{Frontiers2025}.

\textbf{Solution.} In this work, we present a retrieval approach that leverages a Knowledge Graph-based method to effectively capture the hierarchical and relational dependencies inherent in software specifications. Compared to state-of-the-art generic vector-based semantic retrieval and vanilla LLMs (as baseline), ours is
a systematic empirical study that
evaluates energy-aware framework across five diverse datasets using zero-shot, few-shot, and chain-of-thought prompting techniques under accuracy, energy and carbon constraints. Our proposed approach demonstrates a reproducible and environmentally responsible framework for the automated and effective detection of conflict requirements dependency type, comparable to state-of-the-art methods. These results contribute to the advancement of sustainable and automated software requirements dependency detection. We define our research questions with Conflict dependency type detection in focus, as follows:

\begin{enumerate}
    \item [\textbf{RQ1:}] How do Knowledge Graph-based Retrieval (KGR) and Vector-based Semantic Retrieval (VSR) compare across environmental sustainability metrics, specifically energy consumption (Wh), latency (s), and carbon emissions (gCO$_2$eq), relative to retrieval efficiency (Recall@K)? 

  \item [\textbf{RQ2:}] What is the impact of different prompting strategies (Zero-Shot, Few-Shot, and Chain-of-Thought) on the multi-dimensional sustainability profile and classification accuracy (Macro F1-score) of SLMs? 

 \item [\textbf{RQ3:}] To what extent does the integrated retrieval-augmented pipelines improve the sustainability-to-performance ratio compared to vanilla SLM (baseline) inference for requirements classification? 
\end{enumerate}

The main contributions of this research are four fold:
\begin{itemize} [leftmargin=*]
    \item[-] 
    \textit{Empirical comparison:} 
    We present the first systematic empirical comparison of Knowledge-Graph Retrieval (KGR) versus Vector Semantic Retrieval (VSR) under strict energy and carbon constraints for SLM-based requirements dependency detection. While entity-centric GraphRAG ideas exist, their application to energy-aware RE dependency classification with open 7–8B SLMs has not been studied.

    \item[-] \textit{Retrieval Architectures:} We present a novel \textit{KGR} pipeline designed to overcome the structural blindness of traditional \textit{VSR}, capturing the hierarchical dependencies that high-dimensional embeddings often miss. We conduct this empirical evaluation on the five datasets (CDN, OpenCOSS, PURE, UAV, WorldVista, IBM-UAV), small-to-medium in size  which span safety-critical, healthcare, and aerospace domains.
    

    \item[-]  \textit{Sustainability Analysis:} We provide a systematic analysis of how predictive accuracy (F1-score) relates to environmental costs, using a three-pillar sustainability framework that measures energy consumption (Wh), latency (s), and carbon emissions (gCO$_2$eq) across the pipeline.

    \item[-] \textit{Open Science:} We provide an open-source implementation source code, prompts, sensitivity analysis and replication package\footnote{https://anonymous.4open.science/r/Energy-aware-Requirements-Dependency-Classification-99BC/}. 
\end{itemize}
    
    


The rest of the paper is organized as follows. Section \ref{sec:background_sp} presents the preliminary concepts used in this study. Section \ref{sec:methodology} elaborates on our approach and details the methodology. The results are documented in Section \ref{sec:results}, followed by a discussion in Section \ref{sec:discussion}. Related works are documented in Section \ref{sec:related_work} while threats to validity are explained in Section \ref{sec:threats_to_validity}. Section \ref{sec:conc_future_work} concludes the paper. 

\section{Background (Preliminaries)}
\label{sec:background_sp}
In this section, we introduce several key background concepts that are essential for understanding our research contribution. This foundation will help clarify the motivations, methods, and implications of our proposed approach.

\textit{1) Large Language Models (LLMs):} LLMs have recently emerged as powerful tools for semantic reasoning in RE, particularly for identifying duplicate, conflicting, and semantically neutral requirement pairs. Systematic analyses report an increase in the adoption of zero-shot and few-shot prompting strategies for requirement classification tasks, demonstrating strong contextual understanding while reducing manual analysis effort \cite{Zadenoori2025}. These prompting-based approaches enable models to generalize across diverse requirement formulations without extensive task-specific fine-tuning, making them suitable for evolving requirement repositories. Hybrid reasoning approaches further combine language-model inference with structured similarity measures or logical constraints to improve contradiction detection accuracy \cite{Gartner2024,saleem2025passionnet}. Such integrations support more reliable semantic interpretation by balancing probabilistic reasoning with structured decision boundaries, positioning LLMs as an effective foundation for automated requirement relationship modelling.

 \textit{2) Small Language Models (SLMs):} SLMs provide a resource-efficient alternative to large-scale architectures while preserving practical reasoning capabilities. SLMs are generally defined as transformer-based models with parameter sizes of up to or below 13 billion parameters, with many practical deployments operating within the 1–8 billion range. Compared to larger models, SLMs typically require reduced memory, lower computational overhead, and reduced energy consumption during inference-time, while maintaining competitive reasoning performance for domain-specific tasks. Model compression and quantization techniques further enhance this efficiency by reducing numerical precision during inference (e.g., 16-bit, 8-bit, or 4-bit representations), enabling faster execution and lower hardware utilization without substantial degradation in task performance. Such lightweight deployment strategies can significantly reduce energy demand and associated carbon emissions in iterative RE workflows. 

\textit{3) Retrieval-Augmented Generation (RAG):} To enhance contextual grounding and mitigate hallucination, RAG integrates external retrieval mechanisms with language model inference \cite{lewis2020retrieval}. In RE contexts, RAG dynamically retrieves semantically relevant requirements candidates before inference, enabling context-aware classification and improved traceability reasoning \cite{hey_requirements_2025}. This retrieval step helps reduce ambiguity by grounding model predictions in previously specified requirements, thereby supporting incremental requirement analysis workflows. Evaluating retrieval-enhanced architectures under SLM-based inference  enables systematic analysis of performance–carbon trade-offs in relationship modeling while supporting environmentally sustainable automation practices.

\textit{4) Knowledge-Augmented Generation (KAG):} Beyond vector-based retrieval, KAG uses structured knowledge representations, such as entity-centric knowledge graphs, to support relational reasoning across interconnected requirement dependencies \cite{karras2024kgempire,edge2025graphrag}. By combining structured graph traversal with model-based inference, KAG enables systematic and explainable exploration of requirement relationships, facilitating transparency and reproducibility in requirement reasoning pipelines.


\begin{figure*}[!h]
    \centering
\includegraphics[scale=.2]{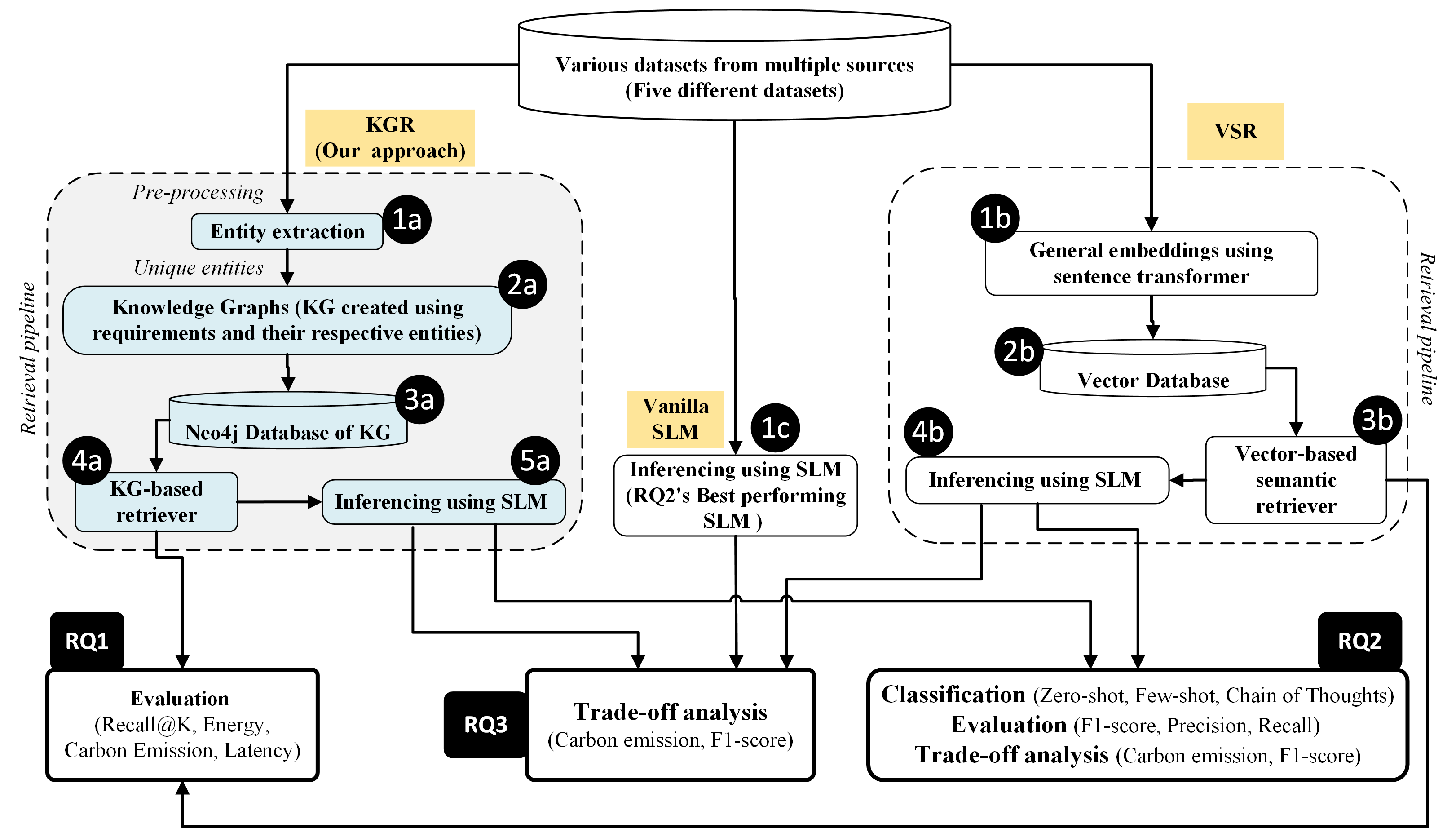}
    \caption{Overall study design: Our proposed KGR approach is compared with VSR and Baseline using five datasets and various evaluation measures.}
\label{fig:framework}
\end{figure*}
\label{sec:methodology}

\section{Methodology}

Our study design is as shown in Figure \ref{fig:framework}. We propose a nuanced information retrieval pipeline (KGR), which is carefully optimized to enable consistent and comparable evaluation across approaches (VSR and baseline) while ensuring the retrieval of contextually relevant requirements for downstream requirements classification task. 

\textit{A. Dataset.}\label{AA}
In this study we have used five software requirements specification (SRS) datasets from heterogeneous domains and requirement styles, such as software, healthcare, transportation, and hardware. Three of these datasets are open-source: OpenCoss, World Vista, and UAV, whereas, PURE and IBM-UAV are extracted from public or proprietary sources. The descriptive information for these is shown in Table \ref{tab:dataset_overview}. The datasets were selected to provide heterogeneous domain coverage, diverse requirement structures, and both open and industrial sources, enabling robust evaluation of generalizability and real-world applicability.

\textit{1) OpenCOSS}~\cite{opencoss_dataset}: A safety-critical system dataset that covers the railway, avionics, and automotive domains. The dataset originates from the OpenCOSS European project and contains requirements focusing on compliance, traceability, and safety assurance.
    
\textit{2) WorldVista}~\cite{worldvista_dataset}: Represents a healthcare information system and captures patient-related processes and clinical workflows. It provides domain-specific requirements characterized by process constraints and contextual dependencies.
    
\textit{3) UAV}\cite{UAVDataset}: Developed at the University of Notre Dame, this dataset uses the EARS template \cite{Mavin2009EARS} for the functional requirements of UAV control systems.

\textit{4) PURE}\cite{PUREDataset}: Consists of a subset of the publicly available PURE dataset, which contains 83 requirements extracted from two software requirements specifications (THEMAS and Mashbot). The dataset includes functional and system-level requirements written in natural language.

\textit{5) IBM-UAV}: A proprietary industrial data set that covers aerospace and automotive domains. Requirements follow IBM’s internal Requirement Quality Analysis (RQA) format, allowing evaluation under realistic industrial constraints

\begin{table}[t]
\centering
\caption{Dataset characteristics for requirement pair datasets, including class distribution, average tokens per pair (as a tuple for both requirements), and vocabulary size (total unique words).}
\label{tab:dataset_overview}
\begin{tabular}{p{1.5cm}p{1cm}p{1cm}p{2cm}p{1cm}}
\textbf{Dataset} & \textbf{\#Conflict (C)} & \textbf{\#Neutral (N)} & \textbf{Avg. \#Tokens in Pairs} & \textbf{Vocabulary Size} \\
\midrule
IBM\_UAV       & 5,553 & 3,400  & (18.19, 18.43) & 1022 \\
WorldVista & 35   & 10,843 & (23.09, 21.80) & 884  \\
UAV        & 18   & 6,652  & (26.44, 26.30) & 298  \\
PURE       & 20   & 2,191  & (26.44, 26.30) & 298  \\
OPENCOSS   & 10   & 6,776  & (29.85, 30.02) & 250  \\
\bottomrule
\end{tabular}
\end{table}

The aggregation of these datasets ensures diversity in domain vocabulary, writing styles, structural rigour, and safety-critical characteristics. This heterogeneity allows us to evaluate the robustness of the retrieval across domains. To avoid redundancy and ensure consistency across experiments, duplicate requirements are removed, resulting in a unified set of unique requirements used for entity extraction and indexing.

\textit{B. Requirement Pair Retrieval.}
\label{sec:information_retrieval}
To allow a fair and structured comparison between retrieval strategies, we design a dual-pipeline information retrieval framework consisting of the \textit{KGR} and \textit{VSR} approaches.


\textit{1) Knowledge Graph-based Retrieval (KGR):} We employ a spaCy-based linguistic pipeline to extract structured entities from each requirement \tikz[baseline=(char.base)]{    \node[shape=circle,draw,fill=black,text=white,inner sep=0pt, minimum size=5pt] (char) {\scriptsize 1a};
    }.
The extraction strategy captures richer semantic structure by identifying domain-specific actors (e.g., organizations or systems), lemmatized action verbs, objects, attributes, and contextual conditions \tikz[baseline=(char.base)]{  \node[shape=circle,draw,fill=black,text=white,inner sep=0pt, minimum size=5pt] (char) {\scriptsize 2a};
    }.

\begin{figure}[!t]
    \centering
\includegraphics[width=\columnwidth]{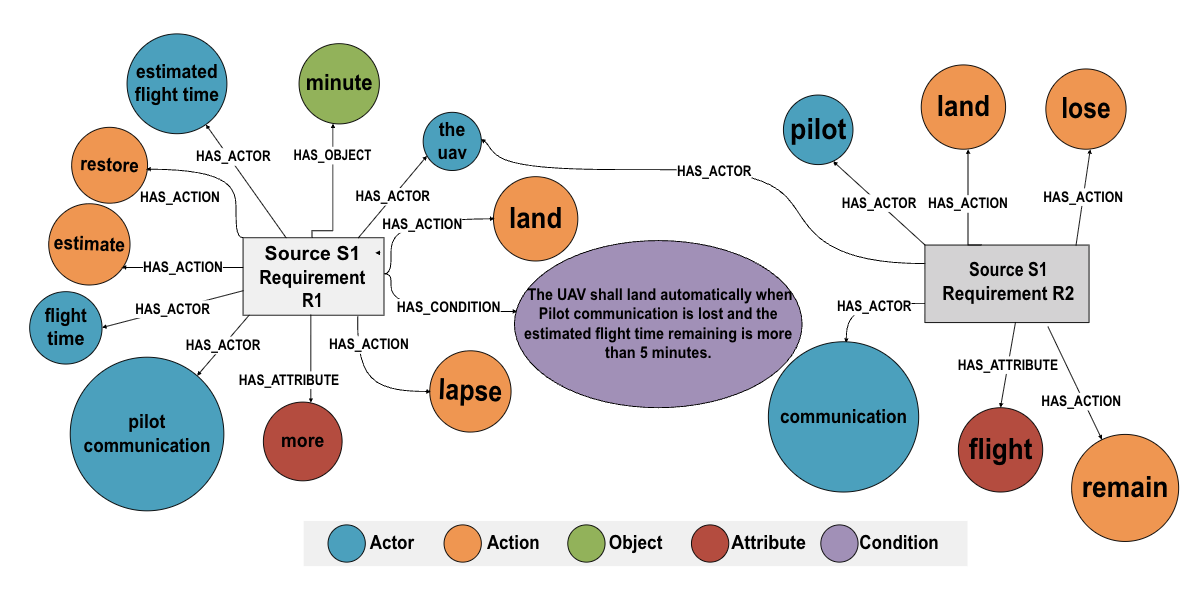}
    \caption{Knowledge graph–based retrieval via shared entities. 
    Two requirements, Requirement R1: \textit{The UAV shall land automatically when Pilot communication is restored and estimated flight time lapsed is more than 5 minutes} and Requirement R2: \textit{The UAV shall land automatically when Pilot communication is lost and the estimated flight time remaining is more than 5 minutes,} are connected through overlapping structured entities, enabling explainable graph traversal.}
    \label{fig:kg_retrieval}
\vspace{-4mm}
\end{figure}

Each requirement and its extracted entities are stored in a Neo4j knowledge graph \tikz[baseline=(char.base)]{    \node[shape=circle,draw,fill=black,text=white,inner sep=0pt, minimum size=5pt] (char) {\scriptsize 3a};
    }. Requirements are represented as nodes, and entity types (Actor, Action, Object, Attribute, Condition) are modeled as distinct node categories. Typed relationships (e.g., HAS\_ACTOR, HAS\_ACTION, HAS\_OBJECT, HAS\_CONDITION, HAS\_ATTRIBUTE) connect requirements to their associated entities. This structured representation enables graph-based semantic retrieval via shared entities \tikz[baseline=(char.base)]{    \node[shape=circle,draw,fill=black,text=white,inner sep=0pt, minimum size=5pt] (char) {\scriptsize 4a};
    }. As illustrated in Figure~\ref{fig:kg_retrieval}, two requirements become connected through shared entities (e.g., actor, object, condition), enabling explainable and multi-hop retrieval. This mechanism allows the system to retrieve semantically related requirements even when surface-level textual similarity is low.

\algrenewcommand\algorithmicensure{\textbf{Output:}}
\begin{algorithm}[!htpb]
\caption{Knowledge-Graph Retrieval  (KGR) via Shared Entities (with $\alpha,\beta,\gamma$ scoring)}
\label{alg:kag_retrieval_alpha_beta_gamma}
\begin{algorithmic}[1]
\Require Query requirement $R_q$, dataset/source id $S$, knowledge graph $G$ (Neo4j), top-$k$ value $k$,  Weights $\alpha,\beta,\gamma$
\Ensure  Ranked list $\mathcal{C}_k$ of top-$k$ retrieved candidate requirements

\State $E_q \gets \mathrm{ExtractEntities}(R_q)$ \Comment{(extraction) Actor/Action/Object/Attribute/Condition}
\State \textbf{if} $E_q = \emptyset$ \textbf{ then return } $\emptyset$
\State $\mathcal{C} \gets \emptyset$
\ForAll{$e \in E_q$}
    \State $\mathcal{C} \gets \mathcal{C}\ \cup\ \{\, R_j \in G.\mathrm{Req}(S)\ :\ R_j \xrightarrow{*} e \,\}$
\EndFor
\State $\mathcal{C} \gets \mathcal{C} \setminus \{R_q\}$

\ForAll{$R_j \in \mathcal{C}$}
    \State $E_j \gets \mathrm{EntitiesOf}(R_j)$
    \State $s_e \gets |E_q \cap E_j|$ \Comment{\# shared entities}
    \State $T_{match} \gets \{\, t\ :\ \exists e \in (E_q \cap E_j)\ \wedge\ (R_q \xrightarrow{t} e)\ \wedge\ (R_j \xrightarrow{t} e)\,\}$
    \State $s_t \gets |T_{match}|$ \Comment{\# matching relationship types}
    \State $d \gets \mathrm{ShortestPathLen}(R_q, R_j)$ \Comment{shortest graph distance}
    \State $s_d \gets \begin{cases}
        \frac{1}{d}, & d>0\\
        0, & \text{otherwise}
    \end{cases}$
    \State $Score[R_j] \gets \alpha\cdot s_e\ +\ \beta\cdot s_t\ +\ \gamma\cdot s_d$
\EndFor

\State \Return $\mathrm{TopK}(\mathcal{C}, Score, k)$
\end{algorithmic}
\end{algorithm}








Algorithm~\ref{alg:kag_retrieval_alpha_beta_gamma} presents our KGR approch implemented in the Neo4jgraph database to identify candidate requirements that are structurally relevant for a given query requirement $R_q$. The algorithm begins by applying the entity extraction process to $R_q$, resulting in a set of structured entities $E_q$. These entities correspond to domain-specific semantic roles such as \textit{Actor}, \textit{Action}, \textit{Object}, \textit{Attribute}, and \textit{Condition}, and serve as the primary anchors for knowledge graph traversal (line 1). Using these extracted entities, the algorithm retrieves candidate requirements by traversing the knowledge graph and collecting all requirements that share at least one entity with $R_q$ (lines 2-7). The union of these requirements forms the candidate set $\mathcal{C}$, excluding the query requirement itself. For each candidate requirement $R_j \in \mathcal{C}$, the algorithm computes a structured similarity score based on three complementary signals defined in the methodological formulation (lines 8-13):
\begin{enumerate}[label=(\roman*)]
    \item First, the entity-overlap component measures the number of shared entities between the query and candidate requirements, represented as $|E_q \cap E_j|$. This term captures direct semantic similarity through common concepts.
    \item Second, the relationship-type agreement component evaluates whether shared entities appear in the same structural role in both requirements. This is quantified by the set $T_{\text{match}}$, whose cardinality reflects the consistency of the semantic roles between the requirements. 
    \item Third, the graph-proximity component incorporates the shortest path distance $d(R_q,R_j)$ between the query and candidate nodes within the knowledge graph. Its reciprocal, $1/d(R_q,R_j)$, assigns greater relevance to candidates that are structurally closer in the graph.
\end{enumerate}

These three components are linearly combined using weighting coefficients $\alpha$, $\beta$, and $\gamma$, resulting in the overall candidate score
\begin{equation}
Score(R_q,R_j)=\alpha |E_q \cap E_j|+\beta |T_{\text{match}}|+\gamma \frac{1}{d(R_q,R_j)}.
\end{equation}
Here, $\alpha$ controls the influence of semantic overlap, $\beta$ emphasizes the consistency of structural roles, and $\gamma$ captures relational proximity within the knowledge graph. By tuning these coefficients, the retrieval mechanism balances lexical similarity, structural alignment, and graph connectivity, we empirically tuned $\alpha$, $\beta$ and $\gamma$ through heuristic experimentation, selecting the combination that produced the best validation performance. Finally, all candidate requirements are ranked according to the computed scores, and the top-$k$ highest-scoring candidates are returned as the retrieval output. This ranked set forms the input for subsequent LLM-based classification, ensuring that downstream reasoning operates over structurally and semantically relevant requirement pairs \tikz[baseline=(char.base)]{    \node[shape=circle,draw,fill=black,text=white,inner sep=0pt, minimum size=5pt] (char) {\scriptsize 5a};
    }. For Pure, UAV, WorldVista, and OpenCoss, entity-overlap-based traversal provides sufficient discrimination due to moderate graph density and limited structural ambiguity. However, the IBM UAV dataset exhibits high entity frequency and structurally rich condition clauses, which necessitate a role-aware, inverse-frequency weighting scheme to mitigate noise from ubiquitous entities and emphasize semantically critical roles. This dataset-specific operationalization preserves the conceptual retrieval framework while improving robustness and computational efficiency.

\textit{2) Vector-based Semantic Retrieval (VSR)} 
In parallel, we construct a semantic retrieval pipeline using dense vector representations. Requirements are embedded using the Sentence-Transformers model \textit{all-mpnet-base-v2}, producing 768-dimensional embeddings \tikz[baseline=(char.base)]{    \node[shape=circle,draw,fill=black,text=white,inner sep=0pt, minimum size=5pt] (char) {\scriptsize 1b};
    }. The embeddings are stored in a Milvus vector database and indexed using IVF\_FLAT with cosine similarity for efficient nearest-neighbor search \cite{farooq2023assessing} \tikz[baseline=(char.base)]{    \node[shape=circle,draw,fill=black,text=white,inner sep=0pt, minimum size=5pt] (char) {\scriptsize 2b};
    }. Such dense vector representations have been shown to be effective for measuring semantic similarity between requirements \cite{Ferrari2017WordEmbeddings}. In particular, prior work such as calculating requirements similarity using word embeddings\cite{reddivari2022calculating} demonstrates that distributed representations capture contextual meaning beyond surface-level lexical overlap. Given a query embedding $v_q$, the system retrieves top-$k$ nearest neighbors based on cosine similarity. This pipeline enables similarity-based retrieval independent of explicit entity overlap and serves as a strong semantic baseline \tikz[baseline=(char.base)]{    \node[shape=circle,draw,fill=black,text=white,inner sep=0pt, minimum size=5pt] (char) {\scriptsize 3b};
    }. This set is further input to LLM-based classification \tikz[baseline=(char.base)]{    \node[shape=circle,draw,fill=black,text=white,inner sep=0pt, minimum size=5pt] (char) {\scriptsize 4b};
    }.
\begin{table}[!t]
\centering
\caption{Open-Source SLMs Used in Our Study (Temperature = 0).}
\label{tab:slms_used}
\begin{tabular}{l c c c}
\hline
\textbf{Model} & \textbf{Params} & \textbf{Release} & \textbf{Context} \\
\hline
DeepSeek-Coder-7B-Instruct-v1.5 & 7B & May 2024 & 16k \\
Meta-Llama-3-8B-Instruct & 8B & Apr 2024 & 8k \\
Mistral-7B-Instruct-v0.3 & 7B & May 2024 & 32k \\
\hline
\end{tabular}
\vspace{-3mm}
\end{table}

\textit{C.  Classification Inference}: We selected open-source SLMs instead of proprietary LLMs in this study as the open-source models enable local deployment, providing full control over execution, monitoring, and instrumentation. This level of control is essential for ensuring transparent energy measurement, accurate carbon accounting, and experimental reproducibility. To evaluate the proposed architecture, we quantify the total energy consumption and carbon emissions of using standalone SLMs for classification \tikz[baseline=(char.base)]{  \node[shape=circle,draw,fill=black,text=white,inner sep=0pt, minimum size=5pt] (char) {\scriptsize 1c};
    } and compare them against our retrieval-augmented selective inferencing framework (\tikz[baseline=(char.base)]{    \node[shape=circle,draw,fill=black,text=white,inner sep=0pt, minimum size=5pt] (char) {\scriptsize 4b};
    } and \tikz[baseline=(char.base)]{    \node[shape=circle,draw,fill=black,text=white,inner sep=0pt, minimum size=5pt] (char) {\scriptsize 5a};
    }). Based on the findings from RQ2, we identify the best-performing configuration i.e., the optimal SLM combined with an effective prompting strategy and use this configuration to measure the overall environmental and performance impact of the system. The selected SLMs used in this study are as listed in Table \ref{tab:slms_used}. The prompts used for classification inferences are shown in the boxes on the right side.

\textit{D. Evaluation metrics.} \label{sec:evaluation_metrics} To systematically assess the proposed architecture, we adopt a multi-objective evaluation framework that measures (i) retrieval efficiency and effectiveness, (ii) classification performance, and (iii) environmental sustainability.

   \begin{tcolorbox}[
  enhanced,
  sharp corners,
  colback=white!98!black,
  colframe=gray!150!black,
  boxrule=0.3pt,
  arc=2pt,
  title={\bfseries Zero-Shot Prompt forConflict Classification},
  fonttitle=\small\bfseries,
  coltitle=black,
  center title,
  top=1pt, bottom=1pt, left=3pt, right=3pt
]

\footnotesize
\textbf{Task context:} You are a requirements analyst.
\newline
\textbf{Goal:} Given an \textbf{ANCHOR} requirement and \textbf{one CANDIDATE} requirement, classify the candidate as:

 - \textbf{Conflicts} with the Anchor  
\newline
 - \textbf{Neutral} to the Anchor
\\
\textbf{Label definitions}

\textbf{Conflict}: Requirements cannot both be true/satisfied simultaneously.  
They impose incompatible or contradictory constraints.

\textbf{Neutral}: Requirements describe different, unrelated or independent behaviour.

\vspace{1mm}
\textbf{ANCHOR:} \texttt{"anchor"}  
\newline
\textbf{CANDIDATE:} \texttt{"candidate"}

\vspace{3mm}
\noindent\bfseries IMPORTANT:\\
Return \textbf{only} the following JSON format — no extra text:

\begin{verbatim}
{
  "label": "Conflict" or "Neutral",
  "confidence": <0–1>
}
\end{verbatim}
\end{tcolorbox}

\textit{1) Recall@K:} Retrieval performance is evaluated using Recall@K, a standard metric in information retrieval for assessing top-K retrieval effectiveness\cite{manning2008ir, baeza1999ir}. 

Let $Q$ denote the set of query requirements, $R_K(q)$ the top candidates retrieved $K$ for query $q$, and $G(q)$ the relevant requirements based on the ground-truth.

For the OpenCOSS, WorldVista, PURE, and UAV datasets, each requirement has exactly one conflicting counterpart. Therefore, Recall@K measures whether the correct conflicting requirement appears within the top-$K$ retrieved candidates:
\begin{equation}
\text{Recall@K} =
\frac{1}{|Q|}
\sum_{q \in Q}
\mathbb{I}\left( R_K(q) \cap G(q) \neq \emptyset \right)
\end{equation}
where $\mathbb{I}(\cdot)$ is the indicator function. In this single-conflict setting, Recall@K reduces to the proportion of queries for which the correct conflicting requirement is retrieved within the top results $K$IQ1.
\begin{tcolorbox}[
  enhanced,
  sharp corners,
  colback=white!98!black,
  colframe=gray!150!black,
  boxrule=0.3pt,
  arc=2pt,
  title={\bfseries Few-Shot Prompt for Conflict Classification},
  fonttitle=\small\bfseries,
  coltitle=black,
  center title,
  top=1pt, bottom=1pt, left=3pt, right=3pt
]
\footnotesize
\textbf{Task context:} You are a requirements analyst.
\newline
\textbf{Goal:} Given an \textbf{ANCHOR} requirement and \textbf{one CANDIDATE} requirement, classify the candidate as:

 - \textbf{Conflicts} with the Anchor  
\newline
 - \textbf{Neutral} to the Anchor
\\
\textbf{Label definitions}
\textbf{Conflict}: Requirements cannot both be true/satisfied simultaneously.  
They impose incompatible or contradictory constraints.

\textbf{Neutral}: Requirements describe different, unrelated or independent behaviour.

\textbf{Examples:}
First Shot, Second Shot, Third Shot

\textbf{ANCHOR:} \texttt{"anchor"}  
\newline
\textbf{CANDIDATE:} \texttt{"candidate"}

\vspace{1mm}
\bfseries MUST return your final answer strictly in the following JSON
format. Do NOT include any text before or after the JSON:
\begin{verbatim}
{
  "label": "Conflict"|"Neutral",
  "confidence": 0–1
}
\end{verbatim}
\end{tcolorbox} 

\begin{tcolorbox}[
  enhanced,
  sharp corners,
  colback=white!98!black,
  colframe=gray!150!black,
  boxrule=0.3pt,
  arc=2pt,
  title={\bfseries COT Prompt for Conflict  Classification},
  fonttitle=\small\bfseries,
  coltitle=black,
  center title,
  top=1pt, bottom=1pt, left=3pt, right=3pt
]
\footnotesize
\textbf{Task context:} You are a requirements analyst.
\newline
\textbf{Goal:} Given an \textbf{ANCHOR} requirement and \textbf{one CANDIDATE} requirement, classify the candidate as:

 - \textbf{Conflicts} with the Anchor  
\newline
 - \textbf{Neutral} to the Anchor
\\
\textbf{Label definitions}

\textbf{Conflict}: Requirements cannot both be true/satisfied simultaneously.  
They impose incompatible or contradictory constraints.

\textbf{Neutral}: Requirements describe different, unrelated or independent behaviour.

\textbf{Thinking guidance (internal only)}  
\newline
1. Identify entities, constraints  
\newline
2. Compare names, quantities, conditions  
\newline
3. Decide conflict / neutral. Do NOT reveal reasoning

\vspace{0.6mm}
\textbf{ANCHOR:} \texttt{"anchor"}  
\newline
\textbf{CANDIDATE:} \texttt{"candidate"}
\vspace{0.8mm}
\newline
\bfseries MUST return your final answer strictly in the following JSON
format. Do NOT include any text before or after the JSON:

\begin{verbatim}
{
  "label": "Conflict" or "Neutral",
  "confidence": <0–1>
}
\end{verbatim}
\end{tcolorbox}




For the IBM-UAV dataset, which involves multi-class requirement relationships rather than single conflicting pairs, $G(q)$ may contain multiple relevant requirements. In this case, Recall@K measures the proportion of relevant requirements retrieved within the top candidates $K$IQ2:

\begin{equation}
\text{Recall@K} =
\frac{1}{|Q|}
\sum_{q \in Q}
\frac{|R_K(q) \cap G(q)|}
{|G(q)|}
\end{equation}

This generalized formulation ensures consistent evaluation across both single-label and multi-label retrieval settings. 

\textit{2) Classification performance:} Results from different prompting strategies and retrieval pipelines are reported using macro-averaged evaluation metrics: Precision (P), Recall (R), and F1-score (F1). Macro-averaged metrics are particularly appropriate for conflict detection tasks, where class distributions may be imbalanced\cite{sokolova2009systematic, opitz2019macro}. For the IBM-UAV dataset, Macro P, R, and F1 are computed across all classes.
d
\textit{3) Sustainability evaluation:} To quantify environmental impact, we instrument all experiments using CodeCarbon\cite{codecarbon}. Energy consumption (kWh) and carbon emissions (kg CO$_2$e) are recorded during inference.

\paragraph{Total Carbon per Dataset per Retrieval Pipeline}

For each dataset $d$ and retrieval pipeline $p$, the total carbon emission is computed as:

\begin{equation}
\text{Carbon}^{(d,p)} =
\sum_{j=1}^{R_{d,p}} C^{\text{retr}}_j
\end{equation}

where $C^{\text{retr}}j$ represents the carbon emission associated with retrieval operation $j$, and $R_{d,p}$ is the total number of retrieval queries executed for dataset $d$ under pipeline $p$. This metric captures the complete environmental cost of executing a specific retrieval pipeline on a given dataset.

\paragraph{EcoScore} To jointly assess predictive performance and sustainability, we define EcoScore at the model level as:

\begin{equation}
\text{EcoScore}_{model} =
\frac{\sum_{d=1}^{D} F1^{(d)}_{macro}}
{\sum_{d=1}^{D} \text{Carbon}^{(d)}}
\end{equation}

where $D$ is the number of datasets, $F1^{(d)}_{macro}$ is the dataset-level Macro F1-score, and $\text{Carbon}^{(d)}$ is the total carbon emission for dataset $d$. EcoScore quantifies predictive performance per unit carbon emission, enabling balanced comparison across models under both effectiveness and environmental efficiency criteria. This formulation is inspired by recent efforts to integrate performance and enviromental cost in AI evaluation \cite{schwartz2020green, patterson2021carbon}.


\textit{E. Hardware/software infrastructure used.}
Retrieval phase (local evaluation) experiments were conducted on a Mac Mini M4 with 16\,GB RAM. Milvus and Neo4j were deployed locally as Docker containers.  Containers were initialized and allowed to reach a stable operational state before measurements began.  Retrieval latency and energy were measured only after index loading and database warm-up, ensuring steady-state evaluation. Inference phase (Cloud + GPU Evaluation) experiments were performed on Google Colab using an NVIDIA A100 GPU with FP16 execution. 
The SLM was loaded once per session. 
Milvus was hosted on Zilliz Cloud, and the knowledge graph was hosted on Neo4j AuraDB. 
During this phase, only the energy and latency of SLM inference (prompt processing and token generation) were tracked. 
Model loading, CUDA initialization, and remote database infrastructure costs were excluded from measurement.

\section{Results}
\label{sec:results}
In this section, we assess the impact and effectiveness of proposed architecture for five different requirement datasets along with the comparison with SLMs\footnote{All the detailed results and per-class Precision/Recall/F1 for Conflict dependency type have been added to our anonymous repository}. 

\subsection{\textit{KGR Vs.VSR - comparison across environmental sustainability metrics (RQ1)}} \noindent We compare Knowledge-Graph-based Retrieval (KGR) and Vector-based Semantic Retrieval (VSR) across energy consumption, carbon emissions, and latency, relative to retrieval effectiveness (Recall@K). The $K$ values used for each dataset were selected based on the elbow point of the Recall@K curve, ensuring that retrieval depth reflects diminishing recall gains while avoiding unnecessary computational overhead. The selected $K$ values are: Pure ($K=3$), OpenCoss ($K=5$), UAV ($K=2$), WorldVista ($K=5$), and IBM\_UAV ($K=20$). 

\noindent\textit{1) Pure Dataset ($K=3$)}:
Both pipelines achieve identical retrieval effectiveness (Recall@3 = 1.00). However, KGR consumes 0.000002 kWh compared to 0.000008 kWh for VSR, representing a 75\% reduction in energy usage. Similarly, carbon emissions decrease from 0.000005 kgCO$_2$e (VSR) to 0.000001 kgCO$_2$e (KGR), corresponding to an 80\% reduction. Latency increases from 380.06 s (VSR) to 555.49 s (KGR), i.e., a 46\% overhead. These results indicate that KGR strictly dominates VSR environmentally while preserving identical recall.

\noindent\textit{2) OpenCoss Dataset ($K=5$):}
At $K=5$, VSR achieves perfect recall (1.00) whereas KGR attains 0.95, resulting in a 5\% absolute difference. Nevertheless, KGR reduces energy consumption by 75\% (0.000003 vs. 0.000012 kWh) and carbon emissions by 71\% (0.000002 vs. 0.000007 kgCO$_2$e). Latency increases by approximately 47\%. Thus, a 5\% recall improvement with VSR requires approximately four times higher environmental cost.

\noindent\textit{3)UAV Dataset ($K=2$):}
Recall values are nearly identical (0.941 for KGR vs. 0.943 for VSR; difference = 0.2\%). Despite this parity, KGR reduces energy consumption by 73\% and carbon emissions by 71\%. Latency overhead remains approximately 47\%. This demonstrates a strong sustainability advantage of KGR with negligible retrieval degradation.

\noindent\textit{4) WorldVista Dataset ($K=5$)}
At $K=5$, VSR achieves Recall@5 = 0.971 compared to 0.957 for KGR (difference = 1.4\%). However, KGR reduces energy consumption by 71\% and carbon emissions by 78\%. Latency increases by approximately 48\%.Hence, a marginal recall gain with VSR incurs a 3--4$\times$ increase in environmental cost.

\noindent\textit{5) IBM\_UAV Dataset ($K=20$)}
IBM\_UAV represents a large-scale, high-depth retrieval scenario. At $K=20$, VSR achieves higher recall (0.995 vs. 0.972; +2.3\%). Additionally, VSR consumes 15\% less energy (0.000869 vs. 0.001026 kWh) and emits 15\% less carbon. However, KGR demonstrates 7\% lower latency. This suggests that vector indexing scales more efficiently in large, dense corpora with high retrieval depth.

\noindent\textbf{Note:} \textit{Cross-Dataset Analysis.} Across four of five datasets (Pure, OpenCoss, UAV, and WorldVista), KGR reduces energy consumption by 71--75\% and carbon emissions by up to 80\%, while maintaining comparable Recall@K (difference $\leq$ 5\%). Although KGR introduces an average latency overhead of approximately 45--48\%, it provides a substantially improved sustainability-to-recall trade-off in small-to-medium datasets.
\begin{tcolorbox}[boxrule=.5pt, sharp corners]
\textbf{RQ1:} Overall, the results indicate that KGR is environmentally advantageous in typical RE corpora, whereas VSR becomes more competitive in high-$K$, large-scale retrieval settings. Only in the large-scale IBM\_UAV dataset, VSR demonstrate superior energy scalability and slightly higher recall, this is due to the additional computation of dense graph traversal.
\end{tcolorbox}

\begin{figure}
    \centering
    \includegraphics[scale=.35]{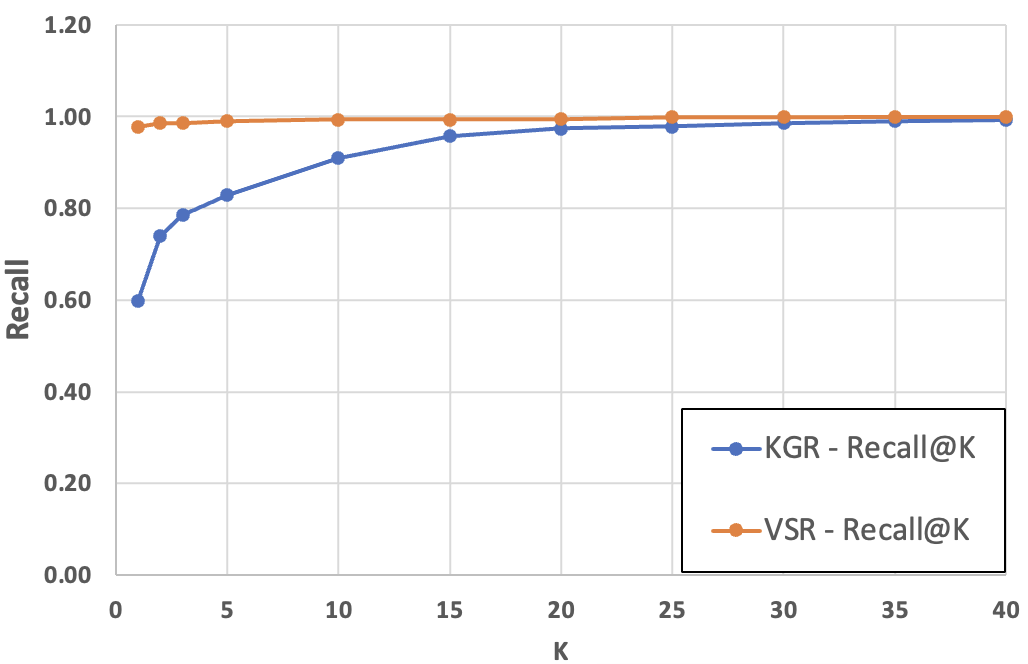}
    \caption{RQ1: Computing K for IBM-UAV performance comparison of KGR and VSR pipelines (Recall@K), the curves plateu at 20. Similar analysis was done for other datasets to choose K value for further evaluation.}
    \label{fig:recallK}
    \vspace{-3mm}
\end{figure}

\begin{figure}
    \centering
    \includegraphics[trim={1cm 1cm .5cm .9cm},clip,scale=.24]{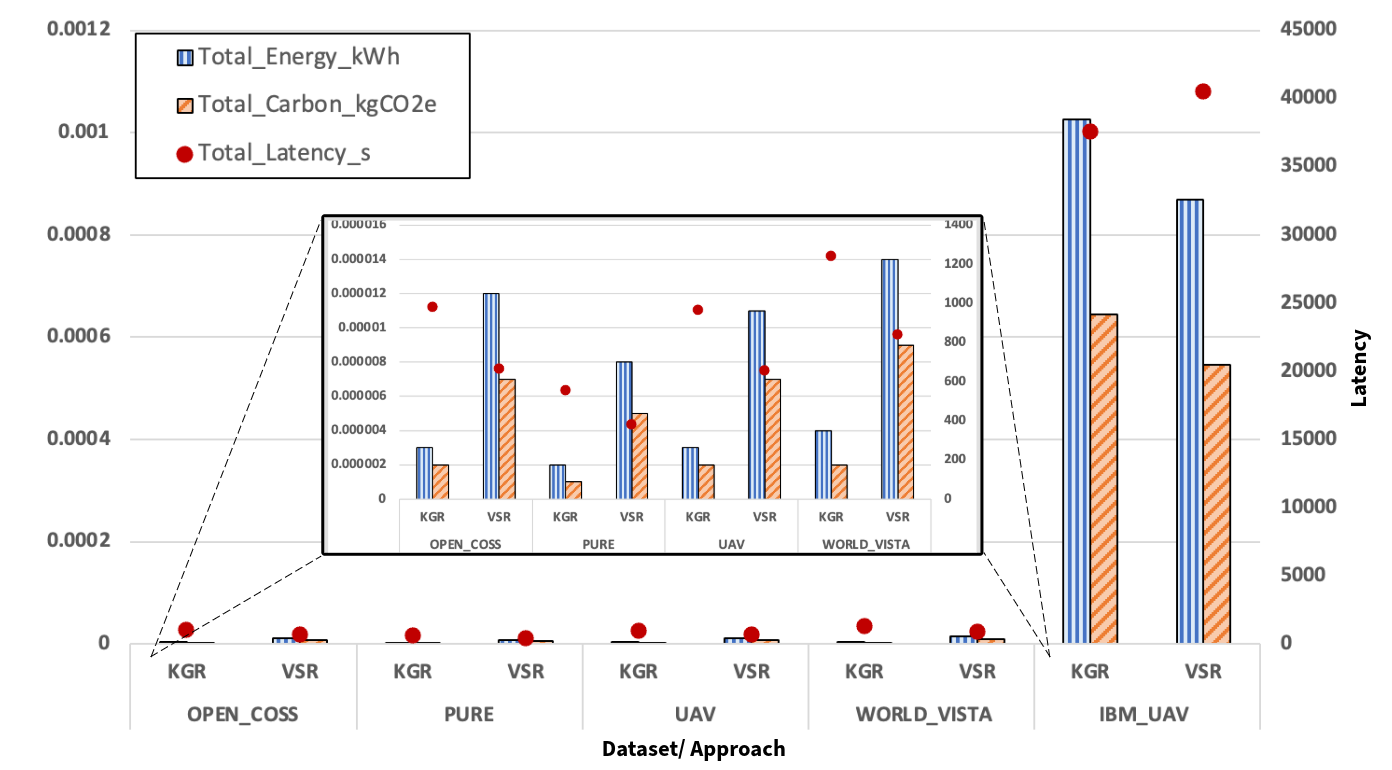}
    \caption{ RQ1 – Comparative evaluation of retrieval pipelines (KGR and VSR) across datasets.}
    \label{fig:RQ1}
    \vspace{-3mm}
\end{figure}






\subsection{\textit{Impact of different prompting strategies on SLMs (RQ2)} }
\begin{figure*}[!htpb]
    \centering
    \includegraphics[scale=.45]{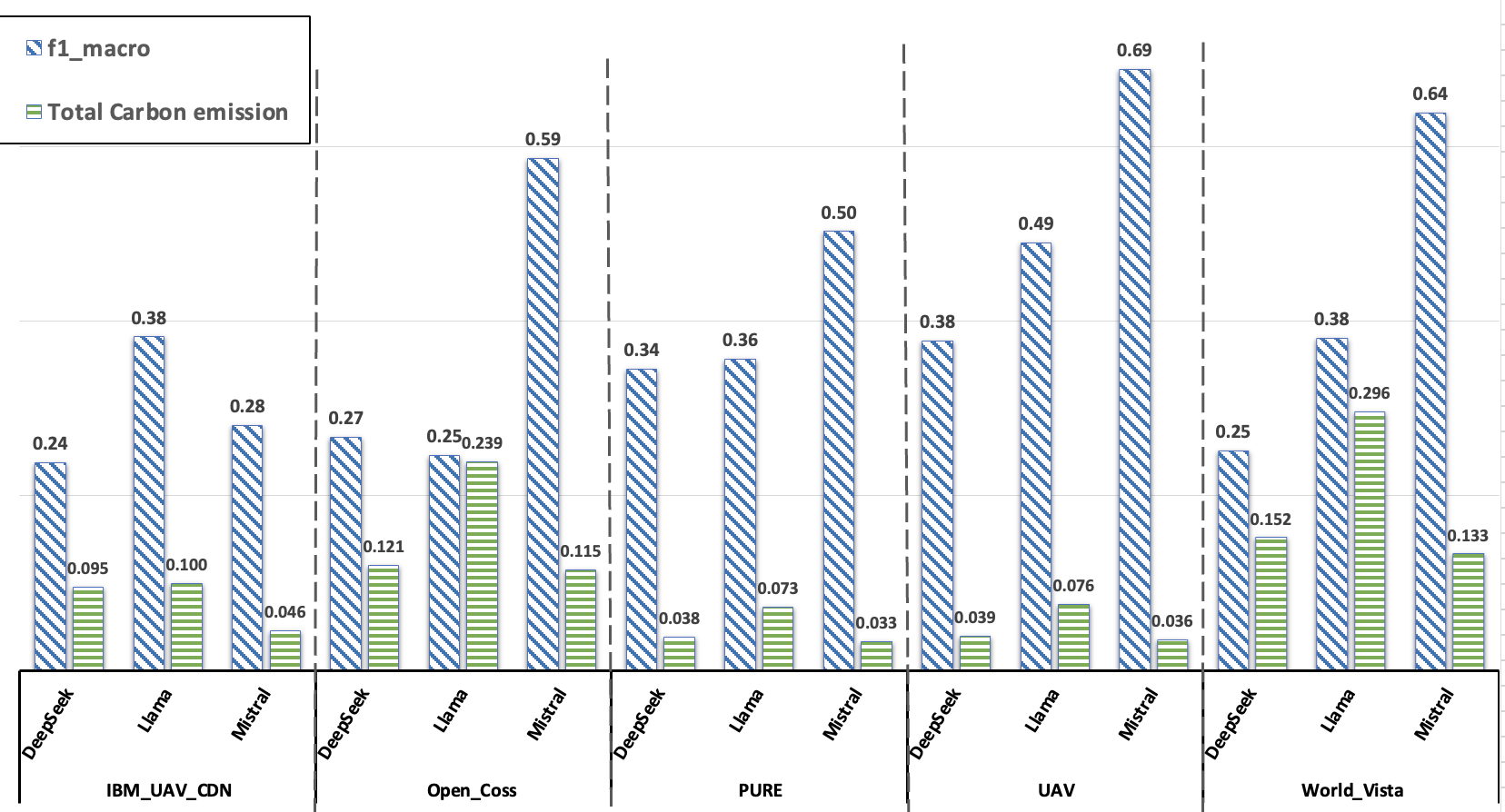}
    \caption{Answering RQ2 - Macro F1-score and Total carbon emission for the three SLMs used for evaluating the five datasets. Mistral SLM fairs well on both for all the datasets}
    \label{fig:RQ2}
    \vspace{-3mm}
\end{figure*}

\noindent To answer RQ2, we analyze the effect of Zero-Shot (ZS), Few-Shot (FS), and Chain-of-Thought (CoT) prompting on Macro F1-score and sustainability metrics (energy, carbon, latency, and tokens) across DeepSeek, Llama, and Mistral models for the five datasets.
\begin{table}[!t]
\centering
\caption{Aggregate Model Sustainability and Performance Summary}
\label{tab:aggregate_model_profile}
\small
\setlength{\tabcolsep}{4pt}
\resizebox{\columnwidth}{!}{
\begin{tabular}{p{1.2cm}p{1cm}p{1cm}p{1cm}p{1cm}p{1cm}}
\hline
Model & Mean F1 & Latency (s) & Mean Energy (kWh) & Total Carbon (kgCO$_2$e) & EcoScore \\
\hline
DeepSeek & 0.3 & 2.17 & $1.44 \times 10^{-4}$ & 0.446 & 0.66\\
Llama    & 0.37 & 2.36 & $1.57 \times 10^{-4}$ & 0.78 & 0.47 \\
\rowcolor{blue!15} Mistral  & 0.539 & 1.04 & $7.25 \times 10^{-5}$ &  0.363 & 1.49 \\
\hline
\end{tabular}
}
\vspace{-5mm}
\end{table}

\textit{1) Aggregate Model-Level Sustainability Profile:} As shown in Table \ref{tab:aggregate_model_profile}, Mistral SLM achieves the highest Macro F1 while consuming the lowest energy and carbon, indicating the strongest overall sustainability–performance balance.
\begin{tcolorbox}[boxrule=0.5pt, sharp corners]
\textbf{RQ2:} Prompting strategies affect both classification accuracy and environmental sustainability. Zero-Shot yields the best sustainability profile by minimizing token usage, energy consumption, and carbon emissions, while still achieving competitive (often superior) Macro F1 for DeepSeek and Mistral. Few-Shot adds substantial environmental cost without consistent accuracy gains. COT delivers large accuracy improvements for Llama but with moderate sustainability penalties. Considering the aggregate sustainability profile (Table~\ref{tab:aggregate_model_profile}), Mistral with Zero-Shot offers the most favorable sustainability–performance trade-off. Thus, prompt design is not just a performance choice but a sustainability decision that shapes the environmental footprint of SLM deployment.
\end{tcolorbox}
\textit{2) Prompting Strategy: Token and Energy Overhead:} Few-Shot prompting nearly doubles token usage relative to Zero-Shot and results in approximately 5--6\% higher carbon emissions. CoT introduces moderate overhead, while Zero-Shot remains the most carbon-efficient strategy. The impact of prompting strategies is strongly model-dependent:
\begin{itemize}[leftmargin=*]
    \item[-] \textit{DeepSeek}: Zero-Shot consistently achieves the highest F1 across most datasets while maintaining the lowest energy and carbon footprint.
    \item[-] \textit{Llama}: Chain-of-Thought significantly improves F1 (e.g., +0.44 absolute gain on UAV compared to Zero-Shot) but increases energy and carbon consumption.
    \item[-] \textit{Mistral}: Zero-Shot or Few-Shot often achieves the best performance, indicating strong inherent reasoning capability without explicit CoT scaffolding.
\end{itemize}
Across all models, Few-Shot prompting incurs the highest token overhead (up to 600 tokens), increasing energy usage by approximately 5--6\% and carbon emissions accordingly, without guaranteed performance gains.

\subsection{\textit{Impact of KGR/VSR-Augmented Pipeline on Sustainability-to-Performance Ratio}} To evaluate the sustainability benefits of the integrated pipeline, we compare it against a projected \textit{exhaustive vanilla inference} baseline. Vanilla inference evaluates all possible candidate requirement pairs without retrieval-based pruning.

\noindent\textbf{Exhaustive Candidate Space:}
From Table~I, the total number of candidate requirement pairs across all datasets is $N_{all}$ = 35,498. After KGR/VSR-based shortlisting, the number of 
$N_{KGR/VSR} = 17,123$. This corresponds to a workload reduction of:

$Reduction = 1 - \frac{N_{KGR/VSR}}{N_{all}} = 51.8\%$

\noindent\textbf{Energy Scaling Analysis:}
Using the best-performing configuration (Mistral model with optimal prompting), the mean per-inference energy consumption is: 

$E_{per} = 7.25 \times 10^{-5} \text{ kWh}$

Projected total energy for exhaustive vanilla inference is: 

$E_{vanilla} = 35,498 \times 7.25 \times 10^{-5} = 2.5736 \text{ kWh}$

Total energy for the KGR/VSR-augmented pipeline is:

$E_{KGR/VSR} = 17,123 \times 7.25 \times 10^{-5} = 1.2414 \text{ kWh}$

\begin{table}[!h]
\vspace{-2mm}
\centering
\caption{Projected Exhaustive Vanilla vs KGR/VSR-Augmented Pipeline (Best Model Configuration)}
\label{tab:rq3_exhaustive}
\begin{tabular}{p{2cm}p{2cm}p{2cm}p{1cm}}
\toprule
\textbf{Pipeline} & \textbf{\# Inferences} & \textbf{Total Energy (kWh)} & \textbf{Reduction} \\
\hline
Baseline & 35,498 & 2.5736 & -- \\
KGR/VSR-Augmented & 17,123 & 1.2414 & 51.8\% \\
\bottomrule
\end{tabular}
\end{table}

\noindent\textbf{Carbon Impact:} Since carbon emissions scale linearly with energy consumption in our measurement framework, total carbon emissions are reduced proportionally by 51.8\% under the KGR/VSR pipeline.

\noindent\textbf{Sustainability-to-Performance Implications:}
Importantly, this computational reduction is achieved while preserving or improving Macro F1-score under the best-performing model configuration. Unlike prompting-based improvements (RQ2), which increase token usage and energy overhead, KGR/VSR improves efficiency by reducing the inference search space.

\begin{tcolorbox}[boxrule=0.5pt, sharp corners]
\textbf{RQ3:}  The integrated KGR-augmented pipeline reduces total computational energy and carbon emissions by approximately 51.8\% compared to exhaustive vanilla SLM inference. By pruning the candidate comparison space prior to model execution, KGR/VSR significantly improves the sustainability-to-performance ratio, demonstrating that structured retrieval offers substantial environmental benefits without sacrificing predictive effectiveness.
\end{tcolorbox}

\section{Discussion}
\label{sec:discussion}
The experimental results demonstrate that model behavior is driven by an interplay between prompting structure and retrieval-based context modulation. This section analyzes the computational and representational logic behind these findings.

\textit{1) Prompting strategies altering sustainability profiles.}
Prompting strategies directly influence energy consumption through token expansion effects. Few-Shot prompting significantly increases input length by injecting exemplars, which forces the transformer to perform more attention operations. Because attention complexity scales quadratically with sequence length, these longer prompts result in a disproportionately higher computational load. Similarly, CoT prompting incurs energy overhead due to extended generation sequences, though it improves accuracy by reducing representation ambiguity through structured reasoning traces. In contrast, Zero-Shot prompting remains the most energy-efficient by minimizing sequence length, though it often fails to provide sufficient guidance for semantically ambiguous requirement pairs.


\textit{2) Improving sustainability-to-performance efficiency with KGR.} 
Knowledge Graph-based Retrieval (KGR) enhances efficiency by modifying the input distribution before it reaches the inference engine. While prompting strategies attempt to improve reasoning by increasing computational effort (token expansion), KGR improves efficiency through search-space contraction. By using graph-based filtering to prune irrelevant requirement pairs prior to model execution, KGR eliminates redundant processing cycles spent on low-signal comparisons. This structural grounding ensures the model processes only high-relevance contexts, achieving environmental benefits without increasing the per-inference energy cost.

\textit{3) Model behavior under structured Context.}
Small Language Models (SLMs) exhibit a particular sensitivity to context structure compared to larger foundation models. Lacking the massive internal representational breadth of their larger counterparts, SLMs benefit significantly from the explicit structural cues provided by KGR. In this framework, the Knowledge Graph acts as an external memory mechanism, compensating for the model's limited latent knowledge. This suggests that for SLMs, sustainability and performance are better optimized through pipeline-level architectural improvements rather than solely increasing prompt complexity.

\textit{4) Implications for sustainable NLP systems.}
Our results indicate that sustainable NLP design should prioritize pre-inference search-space reduction over reasoning verbosity. Effective sustainability gains are achieved by filtering contextual noise and favoring structured retrieval over token-heavy prompting strategies. Future system designs should focus on reducing unnecessary computation at the retrieval stage to ensure that high-accuracy classification remains environmentally viable.

\section{Related work}
\label{sec:related_work}
\noindent\textit{A. RAG in RE:}
\begin{table*}[!ht]
\centering
\caption{Comparison of Related Work in Retrieval, Prompting, and Sustainable Requirement Relationship Analysis}
\label{tab:relatedwork}
\scriptsize
\setlength{\tabcolsep}{4pt}
\resizebox{\textwidth}{!}{
\begin{tabular}{l l l l l l}
\textbf{} & \textbf{Task} & \textbf{Method} & \textbf{Prompt} & \textbf{Dataset} & \textbf{Sustainability} \\
\hline

Zadenoori~\cite{Zadenoori2025} 
& LLMs in RE (SLR) 
& Systematic review 
& Zero/Few (surveyed) 
& Multiple RE datasets 
& No \\

Wang~\cite{wang2024transferre} 
& Conflict/Duplicate 
& Transfer learning 
& Supervised 
& Public RE pairs 
& No \\

Saleem~\cite{saleem2025passionnet} 
& Duplicate + Conflict 
& Hybrid similarity + LLM 
& LLM-based 
& 6 RE benchmarks 
& No \\

Gärtner~\cite{Gartner2024} 
& Contradiction 
& Formal logic + LLM 
& Structured prompts 
& 3 RE datasets 
& No \\

Karras~\cite{karras2024kgempire} 
& RE knowledge modeling 
& Knowledge graph 
& -- 
& 680 RE papers 
& No \\

Edge~\cite{edge2025graphrag} 
& Graph-RAG reasoning 
& Entity graph + retrieval 
& Retrieval-augmented 
& Large corpora 
& Token analysis \\

Hu~\cite{hu2025grag} 
& Graph-RAG QA 
& Subgraph retrieval 
& Hard + Soft 
& WebQSP, ExplaGraphs 
& No \\

Sporsem~\cite{sporsem2025re4rag} 
& RE for RAG 
& Retrieval requirements 
& -- 
& Industrial case 
& No \\

Helmeczi~\cite{helmeczi2023fewshot} 
& Sentence-pair classification 
& Few-shot transformers 
& Few-shot 
& Sentence-pair benchmarks 
& No \\

Huang~\cite{huang2025pe4re} 
& Prompt engineering (RE) 
& SLR 
& Zero/Few/CoT/RAG 
& Multiple RE datasets 
& No \\

White~\cite{white2023promptreview} 
& Prompt patterns 
& Prompt taxonomy 
& CoT / ICL 
& Not dataset-focused 
& No \\

Fu~\cite{fu2025llmco2} 
& Carbon modeling 
& Inference estimation 
& -- 
& LLM runtime traces 
& Carbon only \\

Penzenstadler~\cite{penzenstadler2023sustainre} 
& Sustainability in RE 
& Process integration 
& -- 
& Industrial study 
& Sustainability dimensions \\

Silveira~\cite{silveira2022cresustain} 
& Sustainable RE 
& CRESustain framework 
& -- 
& Academic + industry 
& Sustainability framework \\

van Can~\cite{vancan2025backlogrequirements} 
& Requirement localization 
& SLM-based classification 
& Prompt-based 
& Issue/backlog datasets 
& No \\

Ebrahim~\cite{ebrahim2025slmkalmsre} 
& Requirement enhancement 
& Knowledge-augmented SLM 
& Retrieval prompts 
& Industrial RE corpus 
& No \\

CrossTrace~\cite{crossleveltracingllama} 
& Cross-level traceability 
& Lightweight LM reasoning 
& Few-shot / CoT 
& Traceability datasets 
& No \\
\rowcolor{green!15} \textbf{Our Work} 
& \textbf{Dependency types: Conflict, Neutral}
& \textbf{KGR, VSR, SLM }
& \textbf{Zero, Few, CoT }
& \textbf{5 RE datasets}
& \textbf{Carbon, }\textbf{F1-score trade-off} \\
\end{tabular}
}
\vspace{-3mm}
\end{table*}
Recent research in RE has increasingly explored retrieval-enhanced and hybrid reasoning approaches for modeling requirement relationships. Transfer learning with domain-adapted transformers has demonstrated improved performance in detecting duplicate and conflicting requirement pairs compared to traditional feature-engineering baselines \cite{wang2024transferre}. Hybrid frameworks combining similarity measures with LLM inference further enhance robustness by integrating lexical, semantic, and contextual representations \cite{saleem2025passionnet}. Additionally, logic-guided reasoning frameworks integrate symbolic constraints with language model inference to improve contradiction detection reliability in structured engineering contexts \cite{Gartner2024}.

RAG has been introduced to improve contextual grounding by dynamically retrieving semantically relevant requirement candidates prior to inference \cite{lewis2020retrieval}. In RE-specific settings, retrieval-enhanced traceability and requirement recovery approaches demonstrate that contextual retrieval significantly improves reasoning quality and link validation \cite{hey_requirements_2025,Niu2025}. Beyond vector-based retrieval, knowledge graph infrastructures enable structured modeling of requirement entities and their interrelations \cite{karras2024kgempire}. Graph-based retrieval mechanisms further enhance reasoning by utilizing hierarchical and entity-level contextual structures \cite{edge2025graphrag,hu2025grag}. From an RE process perspective, it has also been argued that RAG systems require explicit “Retrieval Requirements,” as correctness depends on retrieval configuration rather than static model behavior \cite{sporsem2025re4rag}.  While these approaches demonstrate improved accuracy and contextual reasoning, comparative evaluation across retrieval architectures remains limited, and energy-aware assessment is rarely incorporated into retrieval-enhanced RE pipelines.

\textit{B. Sustainability and SLMs in RE:} Recent studies have begun exploring SLMs for RE tasks due to their efficiency and reduced computational footprint. Van Can et al.~\cite{vancan2025backlogrequirements} demonstrate lightweight requirement localization in backlog items, while Ebrahim et al.~\cite{ebrahim2025slmkalmsre} show that SLMs combined with knowledge augmentation can effectively support requirement understanding. Cross-level tracing studies~\cite{crossleveltracingllama} further indicate that fine-tuned smaller models can achieve competitive performance with improved efficiency.

Alongside model size considerations, prompt engineering has emerged as a critical factor in LLM-based RE performance. Systematic reviews categorize zero-shot, few-shot, COT, and retrieval-augmented prompting strategies across requirement tasks, highlighting sensitivity to prompt structure and limited systematic comparison of configurations \cite{huang2025pe4re}. Empirical studies further demonstrate that few-shot calibration improves sentence-pair classification robustness in low-resource settings \cite{helmeczi2023fewshot}. Prompt design patterns and reasoning scaffolds have therefore been recognized as key control mechanisms for LLM behavior in traceability and requirement validation contexts \cite{white2023promptreview,rodriguez2023prompts}.

Despite performance improvements, large-scale transformer models impose significant computational and environmental costs. Prior work emphasizes the substantial carbon footprint associated with deep learning models and transformer-based inference \cite{Strubell2019}. In response, inference-level carbon estimation frameworks enable more accurate modeling of runtime emissions across transformer architectures \cite{fu2025llmco2}. Within RE, sustainability dimensions have been discussed from process and design perspectives, advocating structured integration of environmental, social, and technical considerations during requirements specification and validation \cite{penzenstadler2023sustainre,silveira2022cresustain}.

Prior studies (Table~\ref{tab:relatedwork}) show progress in requirement relationship classification, retrieval-enhanced reasoning, and prompt-based inference for RE tasks, but mostly evaluate retrieval architectures, prompting strategies, and model configurations in isolation, focusing on accuracy. Sustainability is treated largely conceptually, with limited inference-level carbon assessment and little analysis of model scale trade-offs. In particular, the potential of SLMs for energy-efficient, retrieval-enhanced RE reasoning is underexplored. These gaps motivate an integrated framework that jointly examines retrieval design, prompt strategy, model scale, and carbon-aware evaluation for sustainable requirement relationship analysis in evolving software ecosystems. To the best of our knowledge, \textbf{this is the first study} that jointly evaluates retrieval architecture choice, prompting strategy, and full inference-time carbon footprint in RE dependency classification.

\section{Threats to Validity} 
\label{sec:threats_to_validity}
\noindent\textit{a) Internal threats:}
Our results are contingent on the ground-truth reliability of the original datasets; any inherent mislabeling by source creators regarding conflict or neutral status directly impacts classification accuracy. The use of fixed hyperparameters---specifically the retrieval limit $K=10$ and the scaling factors ($\alpha, \beta, \gamma$) for Knowledge Graph composite scoring---may constrain the observed trade-offs between F1-score and environmental sustainability. Also, while the 7B--8B parameter range was selected to prioritize energy efficiency, evaluating larger ($13B+$) or fine-tuned models might yield different accuracy--energy profiles. Finally, to mitigate the non-deterministic nature of SLM outputs, we utilized a majority voting mechanism over three prompt executions per instance to ensure more stable and reproducible results.

\noindent \textit{b) External threats:} Our evaluation utilized five English-language datasets from safety-critical, healthcare, and embedded domains; consequently, the performance of these methods on non-English requirements or larger-scale repositories remains a subject for future work. Our selection of three 7B--8B parameter models was intended to balance consumer-grade hardware compatibility with semantic reasoning depth, though sustainability profiles may vary across different architectures. To ensure measurement reliability, we mitigated cold-start bias by excluding initialization transients---such as model loading and CUDA kernel compilation---from our energy tracking. By strictly isolating energy consumption to the execution phase, our methodology reflects operational runtime behavior in alignment with established benchmarking principles~\cite{hennessy_patterson}, experimental rigor guidelines in SE~\cite{cruz_guidelines}, and the transparency standards of Green AI~\cite{schwartz_green_ai}.  Also, we did not compare against fine-tuned models or commercial LLMs (e.g., GPT-4o-mini) because the goal was reproducible open SLM evaluation.

\noindent \textit{c) Construct threats:}
Our \textit{EcoScore} computation accounts exclusively for inference-time energy (prompt processing and token generation), which may omit peripheral system overheads. Again, while we rely on established ground-truth labels to mitigate classification risks, we did not perform manual human validation of the retrieved candidates. This limitation implies that subtle nuances in the requirements, which might influence the classification but are not captured by quantitative metrics, could remain undetected.


\section{Conclusion and Future work}
\label{sec:conc_future_work}
Motivated by the increasing need for sustainable automation approaches in RE, in this study, we proposed an energy-sensitive framework for automated requirements dependency classification and empirically evaluated it on diverse datasets. Also, systematically compared Knowledge-Graph-based Retrieval (KGR) with conventional Vector-based Semantic Retrieval (VSR) and baseline SLMs when augmenting Small Language Models (7B–8B) for Conflict vs. Neutral dependency detection. Across five heterogeneous datasets and three prompting strategies, our proposed KGR approach delivered lower energy consumption and carbon emissions than VSR and the baseline. These results demonstrate that structured graph traversal outperforms semantic vector search for sustainable, explainable requirement dependency detection and establish a reproducible blueprint for green AI in Requirements Engineering. 
In the future, we will evaluate our proposed KGR on larger industrial datasets and more SLM families with human-in-the-loop validation, focusing on full life-cycle carbon accounting.


\bibliographystyle{ieeetr}
\bibliography{references}

@book{Sommerville2016,
  author    = {Ian Sommerville},
  title     = {Software Engineering},
  edition   = {10},
  publisher = {Pearson},
  year      = {2016}
}

@article{Zadenoori2025,
  author    = {Mohammad Amin Zadenoori and Jacek D{\k{a}}browski and Waad Alhoshan and Liping Zhao and Alessio Ferrari},
  title     = {Large Language Models (LLMs) for Requirements Engineering (RE): A Systematic Literature Review},
  journal   = {arXiv preprint},
  year      = {2025},
  archivePrefix = {arXiv},
  eprint    = {2509.11446}
}

@article{cheng2026generative,
  title={Generative ai for requirements engineering: A systematic literature review},
  author={Cheng, Haowei and Husen, Jati H and Lu, Yijun and Racharak, Teeradaj and Yoshioka, Nobukazu and Ubayashi, Naoyasu and Washizaki, Hironori},
  journal={Software: Practice and Experience},
  volume={56},
  number={2},
  pages={141--170},
  year={2026},
  publisher={Wiley Online Library}
}

@misc{HowtoHan57:online,
author = {Anushtha Jain},
  title = {How to Handle Conflicting Requirements in Business Systems - Visure Solutions},
  url = "https://visuresolutions.com/alm-guide/conflicting-requirements/",
month = {9},
year = {2025},
  note = "[Online; accessed 2026-02-26]"
}

@article{nuseibeh2002leveraging,
  title={Leveraging inconsistency in software development},
  author={Nuseibeh, Bashar and Easterbrook, Steve and Russo, Alessandra},
  journal={Computer},
  volume={33},
  number={4},
  pages={24--29},
  year={2002},
  publisher={IEEE}
}

@article{gharib2025reinta,
  title={ReInTa: A Novel Requirements Interdependencies Taxonomy},
  author={Gharib, Mohamad and Mirzazada, Elvin},
  journal={Baltic Journal of Modern Computing},
  volume={13},
  number={4},
  pages={834--861},
  year={2025},
  publisher={University of Latvia}
}

@article{dahlstedt2005requirements,
  title={Requirements interdependencies: state of the art and future challenges},
  author={Dahlstedt, {\AA}sa G and Persson, Anne},
  journal={Engineering and managing software requirements},
  pages={95--116},
  year={2005},
  publisher={Springer}
}

@inproceedings{hey_requirements_2025,
  author    = {Tobias Hey and Dominik Fuch{\ss} and Jan Keim and Anne Koziolek},
  title     = {Requirements Traceability Link Recovery via Retrieval-Augmented Generation},
  booktitle = {Proceedings of the 31st International Working Conference on Requirements Engineering: Foundation for Software Quality (REFSQ 2025)},
  year      = {2025},
  publisher = {Springer},
  doi       = {10.1007/978-3-031-88531-0_27}
}

@article{Niu2025,
  author    = {Feifei Niu and Rongqi Pan and Lionel C. Briand and Hanyang Hu and Krishna Koravadi},
  title     = {TVR: Automotive System Requirement Traceability Validation and Recovery Through Retrieval-Augmented Generation},
  journal   = {arXiv preprint},
  year      = {2025},
  archivePrefix = {arXiv},
  eprint    = {2504.15427}
}

@article{Klesel2025,
  author    = {Michael Klesel and H. Felix Wittmann},
  title     = {Retrieval-Augmented Generation (RAG)},
  journal   = {Business \& Information Systems Engineering},
  year      = {2025},
  volume    = {67},
  pages     = {551--561}
}

@article{Norheim2024,
  author    = {Johannes J. Norheim and Eric Rebentisch and Dekai Xiao and Lorenz Draeger and Alain Kerbrat and Olivier L. de Weck},
  title     = {Challenges in Applying Large Language Models to Requirements Engineering Tasks},
  journal   = {Design Science},
  year      = {2024},
  doi       = {10.1017/dsj.2024.8}
}

@article{Strubell2019,
  author    = {Emma Strubell and Ananya Ganesh and Andrew McCallum},
  title     = {Energy and Policy Considerations for Deep Learning in NLP},
  journal   = {Proceedings of the 57th Annual Meeting of the Association for Computational Linguistics},
  year      = {2019},
  pages     = {3645--3650}
}

@article{lewis2020retrieval,
  title={Retrieval-augmented generation for knowledge-intensive nlp tasks},
  author={Lewis, Patrick and Perez, Ethan and Piktus, Aleksandra and Petroni, Fabio and Karpukhin, Vladimir and Goyal, Naman and K{\"u}ttler, Heinrich and Lewis, Mike and Yih, Wen-tau and Rockt{\"a}schel, Tim and others},
  journal={Advances in neural information processing systems},
  volume={33},
  pages={9459--9474},
  year={2020}
}

@article{Gartner2024,
  author = {Gärtner, Alexander Elenga and Göhlich, Dietmar},
  title = {Automated requirement contradiction detection through formal logic and LLMs},
  journal = {Automated Software Engineering},
  year = {2024},
  volume = {31},
  number = {2},
  pages = {49},
  doi = {10.1007/s10515-024-00452-x},
  url = {https://doi.org/10.1007/s10515-024-00452-x}
}

@article{Frontiers2025,
  author = {Authors, Multiple},
  title = {Advancing engineering research through context-aware and knowledge graph–based retrieval-augmented generation},
  journal = {Frontiers in Artificial Intelligence},
  year = {2025},
  volume = {8},
  doi = {10.3389/frai.2025.XXXXX}
}

@inproceedings{rodriguez2023prompts,
  title={Prompts matter: Insights and strategies for prompt engineering in automated software traceability},
  author={Rodriguez, Alberto D and Dearstyne, Katherine R and Cleland-Huang, Jane},
  booktitle={2023 IEEE 31st International Requirements Engineering Conference Workshops (REW)},
  pages={455--464},
  year={2023},
  organization={IEEE}
}

@article{saleem2025passionnet,
  author  = {Summra Saleem and Muhammad Nabeel Asim and Andreas Dengel},
  title   = {PassionNet: An Innovative Framework for Duplicate and Conflicting Requirements Identification},
  journal = {Expert Systems With Applications},
  volume  = {293},
  pages   = {128684},
  year    = {2025},
  doi     = {10.1016/j.eswa.2025.128684}
}

@article{karras2024kgempire,
  author  = {Oliver Karras},
  title   = {KG-EmpiRE: A Community-Maintainable Knowledge Graph for a Sustainable Literature Review on the State and Evolution of Empirical Research in Requirements Engineering},
  journal = {arXiv preprint arXiv:2405.08351},
  year    = {2024},
  url     = {https://arxiv.org/abs/2405.08351}
}

@article{edge2025graphrag,
  author  = {Darren Edge and Ha Trinh and Newman Cheng and Joshua Bradley and Alex Chao and Apurva Mody and Steven Truitt and Dasha Metropolitansky and Robert Osazuwa Ness and Jonathan Larson},
  title   = {From Local to Global: A GraphRAG Approach to Query-Focused Summarization},
  journal = {arXiv preprint arXiv:2404.16130},
  year    = {2025},
  note    = {Under review},
  url     = {https://arxiv.org/abs/2404.16130}
}

@article{hu2025grag,
  author  = {Yuntong Hu and Zhihan Lei and Zheng Zhang and Bo Pan and Chen Ling and Liang Zhao},
  title   = {GRAG: Graph Retrieval-Augmented Generation},
  journal = {arXiv preprint arXiv:2405.16506},
  year    = {2025},
  url     = {https://arxiv.org/abs/2405.16506}
}

@inproceedings{sporsem2025re4rag,
  author    = {Tor Sporsem and Rasmus Ulfsnes},
  title     = {Towards Requirements Engineering for RAG Systems},
  booktitle = {Proceedings of the International Conference on Evaluation and Assessment in Software Engineering (EASE)},
  year      = {2025},
  address   = {Istanbul, Turkey},
  note      = {arXiv:2505.07553},
  url       = {https://arxiv.org/abs/2505.07553}
}

@article{helmeczi2023fewshot,
  author  = {Péter Helmeczi and Zoltán Istenes and Gábor Berend},
  title   = {Few-Shot Learning for Sentence Pair Classification},
  journal = {arXiv preprint arXiv:2306.08058},
  year    = {2023},
  url     = {https://arxiv.org/abs/2306.08058}
}

@article{huang2025pe4re,
  author  = {Yuxuan Huang and Alessandro Ferrari and Liping Zhao and others},
  title   = {Prompt Engineering for Requirements Engineering: A Systematic Literature Review and Research Roadmap},
  journal = {arXiv preprint arXiv:2507.07682},
  year    = {2025},
  url     = {https://arxiv.org/abs/2507.07682}
}

@article{white2023promptreview,
  author  = {Jules White and Quchen Fu and Sam Hays and Michael Sandborn and Carlos Escobar and others},
  title   = {A Prompt Pattern Catalog to Enhance Prompt Engineering with ChatGPT},
  journal = {arXiv preprint arXiv:2302.11382},
  year    = {2023},
  url     = {https://arxiv.org/abs/2302.11382}
}

@inproceedings{fu2025llmco2,
  author    = {Yiming Fu and Zhenyu Zhou and Yuxuan Zhang and others},
  title     = {LLMCO2: Accurate Carbon Footprint Prediction for LLM Inference},
  booktitle = {Proceedings of HotCarbon 2025},
  year      = {2025},
  note      = {Preprint available at arXiv:2410.02950},
  url       = {https://arxiv.org/abs/2410.02950}
}

@article{wang2024transferre,
  author  = {Xinyu Wang and Jianjun Zhao and others},
  title   = {Transfer Learning for Conflict and Duplicate Detection in Software Requirement Pairs},
  journal = {Journal of Systems and Software},
  year    = {2024},
  note    = {Preprint version analyzed},
  url     = {<insert DOI or official publication link if available>}
}

@article{penzenstadler2023sustainre,
  author  = {Birgit Penzenstadler and Stefanie Betz and Norbert Seyff and others},
  title   = {Sustainability Design in Requirements Engineering: State of Practice},
  journal = {Requirements Engineering},
  year    = {2023},
  publisher = {Springer},
  doi     = {<insert DOI if available>}
}

@article{silveira2022cresustain,
  author  = {Clara Silveira and Vitor Santos and Leonilde Reis and Henrique Mamede},
  title   = {CRESustain: Approach to Include Sustainability and Creativity in Requirements Engineering},
  journal = {Journal of Engineering Research and Sciences},
  volume  = {1},
  number  = {8},
  pages   = {27--34},
  year    = {2022},
  doi     = {10.55708/js0108004}
}

@inproceedings{Ferrari2017WordEmbeddings,
  author    = {Alessio Ferrari and others},
  title     = {Calculating Requirements Similarity Using Word Embeddings},
  booktitle = {Proceedings of the International Working Conference on Requirements Engineering: Foundation for Software Quality (REFSQ)},
  year      = {2017}
}

@inproceedings{reddivari2022calculating,
  title={Calculating requirements similarity using word embeddings},
  author={Reddivari, Sandeep and Wolbert, Jeffery},
  booktitle={2022 IEEE 46th Annual Computers, Software, and Applications Conference (COMPSAC)},
  pages={438--439},
  year={2022},
  organization={IEEE}
}

@inproceedings{Mavin2009EARS,
  author    = {Alistair Mavin and Philip Wilkinson and Adrian Harwood and Mark Novak},
  title     = {Easy Approach to Requirements Syntax (EARS)},
  booktitle = {IEEE International Requirements Engineering Conference},
  year      = {2009}
}

@inproceedings{PUREDataset,
  title={Pure: A dataset of public requirements documents},
  author={Ferrari, Alessio and Spagnolo, Giorgio Oronzo and Gnesi, Stefania},
  booktitle={2017 IEEE 25th international requirements engineering conference (RE)},
  pages={502--505},
  year={2017},
  organization={IEEE}
}

@misc{UAVDataset,
  title        = {UAV Requirements Dataset, University of Notre Dame},
  year         = {2014}
}

@article{vancan2025backlogrequirements,
  title   = {Locating requirements in backlog items: Content analysis and experiments with language models},
  author  = {A. T. van Can and others},
  journal = {Information and Software Technology},
  year    = {2025},
  url     = {https://www.sciencedirect.com/science/article/pii/S0950584924002490}
}

@inproceedings{ebrahim2025slmkalmsre,
  title     = {Enhancing Software Requirements Engineering with Small Language Models and Knowledge-Augmented Language Models},
  author    = {M. Ebrahim and others},
  booktitle = {Proceedings of the ACL Student Research Workshop},
  year      = {2025},
  url       = {https://aclanthology.org/2025.acl-srw.31.pdf}
}

@misc{crossleveltracingllama,
  title  = {Cross-Level Requirements Tracing Based on Large Language Models},
  author = {Authors, Unknown},
  year   = {2024},
  note   = {Compares fine-tuning strategies across LLaMA scales (1.1B, 7B, 13B)},
  url    = {https://www.researchgate.net/publication/391975387_Cross-Level_Requirements_Tracing_Based_on_Large_Language_Models}
}

@book{hennessy_patterson,
  author    = {John L. Hennessy and David A. Patterson},
  title     = {Computer Architecture: A Quantitative Approach},
  publisher = {Morgan Kaufmann},
  year      = {2017}
}

@article{schwartz_green_ai,
  author  = {Roy Schwartz and Jesse Dodge and Noah A. Smith and Oren Etzioni},
  title   = {Green AI},
  journal = {Communications of the ACM},
  year    = {2020}
}

@article{cruz_guidelines,
  author  = {Jesús Cruz-Lemus and others},
  title   = {Experimental Guidelines for Software Engineering Research},
  journal = {Empirical Software Engineering},
  year    = {2019}
}

@misc{opencoss_dataset,
  title        = {OpenCOSS Project},
  howpublished = {\url{http://www.opencoss-project.eu}},
  note         = {Accessed: 2026-02-24}
}

@misc{worldvista_dataset,
  title        = {WorldVista Dataset},
  howpublished = {\url{http://coest.org/datasets}},
  note         = {Accessed: 2026-02-24}
}

@article{farooq2023assessing,
  title={Assessing similarity between software requirements: A semantic approach},
  author={Farooq, Ahmad and Faisal, Mohammad},
  journal={International Journal of Information Engineering and Electronic Business},
  volume={10},
  number={2},
  pages={38},
  year={2023},
  publisher={Modern Education and Computer Science Press}
}

@book{manning2008ir,
  title={Introduction to Information Retrieval},
  author={Manning, Christopher D. and Raghavan, Prabhakar and Schütze, Hinrich},
  year={2008},
  publisher={Cambridge University Press}
}

@book{baeza1999ir,
  title={Modern Information Retrieval},
  author={Baeza-Yates, Ricardo and Ribeiro-Neto, Berthier},
  year={1999},
  publisher={Addison-Wesley}
}

@article{sokolova2009systematic,
  title={A systematic analysis of performance measures for classification tasks},
  author={Sokolova, Marina and Lapalme, Guy},
  journal={Information Processing \& Management},
  volume={45},
  number={4},
  pages={427--437},
  year={2009}
}

@article{opitz2019macro,
  title={Macro F1 and Macro F1},
  author={Opitz, Juri and Burst, Sebastian},
  journal={arXiv preprint arXiv:1911.03347},
  year={2019}
}

@article{schwartz2020green,
  title={Green AI},
  author={Schwartz, Roy and Dodge, Jesse and Smith, Noah A. and Etzioni, Oren},
  journal={Communications of the ACM},
  volume={63},
  number={12},
  pages={54--63},
  year={2020}
}

@article{patterson2021carbon,
  title={Carbon Emissions and Large Neural Network Training},
  author={Patterson, David and Gonzalez, Joseph and Le, Quoc and others},
  journal={arXiv preprint arXiv:2104.10350},
  year={2021}
}

@misc{codecarbon,
  title = {CodeCarbon: A Tool to Estimate Carbon Emissions from Computing},
  author = {Lacoste, Alexandre and other contributors},
  howpublished = {\url{https://github.com/mlco2/codecarbon}},
  year = {2021}
}

\end{document}